\documentclass[journal,11pt,onecolumn,draftclsnofoot]{IEEEtran}
\ifCLASSINFOpdf
\else
\fi

\usepackage{amsfonts}
\usepackage{graphicx}
\usepackage{subfigure}
\usepackage{color}
\usepackage{amssymb}
\usepackage{amsmath}
\usepackage{amsbsy}
\usepackage{indentfirst}
\usepackage{cite}

\newtheorem{theorem}{Theorem}

\newtheorem{definition}{Definition}
\newtheorem{proposition}{Proposition}
\newtheorem{lemma}{Lemma}

\usepackage[font=small]{caption}

\begin{document}
	%
	\title{Sparse recovery based on $q$-ratio constrained minimal singular values}
	%
	%
	%
	
	\author{Zhiyong~Zhou 
		and~Jun~Yu 
		\thanks{The authors are with the Department
			of Mathematics and Mathematical Statistics, Ume{\aa} University,  Ume{\aa},
			901 87, Sweden (e-mail: zhiyong.zhou@umu.se, jun.yu@umu.se).}}
	\maketitle
	
	\begin{abstract}
		We study verifiable sufficient conditions and computable performance bounds for sparse recovery algorithms such as the Basis Pursuit, the Dantzig selector and the Lasso estimator, in terms of a newly defined family of quality measures for the measurement matrices. With high probability, the developed measures for subgaussian random matrices are bounded away from zero as long as the number of measurements is reasonably large. Comparing to the restricted isotropic constant based performance analysis, the arguments in this paper are much more concise and the obtained bounds are tighter. Numerical experiments are presented to illustrate our theoretical results.
	\end{abstract}
	

	%
	\IEEEpeerreviewmaketitle

	\section{Introduction}
	%
	%
	%
	%
	
	\IEEEPARstart{S}{parse} signal recovery, particularly compressive sensing \cite{crt,d,ek,fr}, aims to reconstruct a sparse signal from noisy underdetermined linear measurements: \begin{align}
	y=Ax+w,
	\end{align}
	where $x\in\mathbb{R}^N$ is the true sparse or compressible signal, $y\in\mathbb{R}^m$ is the measurement vector with $m\ll N$, $A\in\mathbb{R}^{m\times N}$ is the measurement matrix, and $w\in\mathbb{R}^m$ is the noise vector. If the measurement matrix satisfies the stable or robust null space property (NSP) \cite{cdd} or restricted isometry property (RIP) \cite{crt,c}, stable and robust recovery can be guaranteed. Although probabilistic results conclude that the NSP and RIP are fulfilled for some specific random matrices with high probability \cite{bddw,rv,sxh}, it's computationally hard to verify NSP and compute restricted isometry constant (RIC) for a given measurement matrix \cite{bdms,tp}. Several relaxation techniques are used to obtain an approximate solution, for instance semi-definite programming \cite{de,dgjl} and linear programming \cite{jn}. Recently, \cite{tn1} and \cite{tn3} defined new kinds of computable quality measures of the measurement matrices. Specifically, \cite{tn1} developed $\ell_1$-constrained minimal singular values (CMSV) $\rho_s(A)=\min\limits_{z\neq 0, \lVert z\rVert_1^2/\lVert z\rVert_2^2\leq s}\frac{\lVert Az\rVert_2}{\lVert z\rVert_2}$ and obtained the error $\ell_2$ bounds in terms of this quality measure of the measurement matrix. Similarly, in \cite{tn3}, the authors defined another quantity $\omega_{\lozenge}(A,s)=\min\limits_{z\neq 0, \lVert z\rVert_1/\lVert z\rVert_\infty\leq s}\frac{\lVert Az\rVert_{\lozenge}}{\lVert z\rVert_\infty}$ with $\lVert\cdot\rVert_{\lozenge}$ denoting a general norm, and derived the performance bounds on the $\ell_{\infty}$ norm of the recovery error vector based on this quality measure. This kind of measures has also been used in establishing results for block sparsity recovery \cite{tn4} and low-rank matrix recovery \cite{tn2}. In this paper we generalize these two quantities to a more general quantity called $q$-ratio CMSV with $1<q\leq \infty $, and establish the performance bounds for both $\ell_q$ norm and $\ell_1$ norm of the reconstruction error.

	\subsection{Contributions}
	Our contribution mainly has four aspects. First, we proposed a sufficient condition based on a $q$-ratio sparsity level for the exact recovery using $\ell_1$ minimization in the noise free case, and designed a convex-concave procedure to solve the corresponding non-convex problem, leading to an acceptable verification algorithm. Second, we introduced $q$-ratio CMSV and derived concise bounds on both $\ell_q$ norm and $\ell_1$ norm of the reconstruction error for the Basis Pursuit (BP) \cite{cds}, the Dantzig selector (DS) \cite{ct}, and the Lasso estimator \cite{t} in terms of $q$-ratio CMSV.  We established the corresponding stable and robust recovery results involving both sparsity defect and measurement error. Third, we demonstrated that for subgaussion random matrices, the $q$-ratio CMSVs are bounded away from zero with high probability, as long as the number of measurement is large enough. Finally, we presented algorithms to compute the $q$-ratio CMSV for an arbitrary measurement matrix, and studied the effects of different parameters on the proposed $q$-ratio CMSV. Moreover, we illustrated that $q$-ratio CMSV based bound is tighter than the RIC based one.
	
	\subsection{Organization and Notations}
	
	The paper is organized as follows. In Section II, we present the definitions of $q$-ratio sparsity and $q$-ratio CMSV, and give a sufficient condition for unique noiseless recovery based on the $q$-ratio sparsity and an inequality for the $q$-ratio CMSV. In Section III, we derive performance bounds on both $\ell_q$ norm and $\ell_1$ norm of the reconstruction errors for several convex recovery algorithms in terms of $q$-ratio CMSVs. In Section IV, we demonstrate that the subgaussian random matrices have non-degenerate $q$-ratio CMSVs with high probability as long as the number of measurements is relatively large. In Section V, we design algorithms to verify the sufficient condition for unique recovery in noise free case and compute the $q$-ratio CMSV. Section VI contains the conclusion. Finally, the proofs are postponed to the Appendix.
	
	Throughout the paper, we denote vectors by lower case letters and matrices by upper case letters. Vectors are columns by default. $z^T$ denotes the transpose of the vector $z$ and $z_i$ denotes the $i$-th entry of $z$. For any vector $z\in\mathbb{R}^N$, we denote the $\ell_0$ norm $\lVert z\rVert_0=\sum_{i=1}^N 1\{z_i\neq 0\}$, the $\ell_{\infty}$ norm $\lVert z\rVert_{\infty}=\max_{1\leq i\leq N}|z_i|$ and the $\ell_q$ norm $\lVert z\rVert_{q}=(\sum_{i=1}^N |z_i|^q)^{1/q}$ for $0<q<\infty$. We say a signal $x$ is $k$-sparse if $\lVert x\rVert_0\leq k$. $[N]$ denotes the set $\{1,2,\cdots,N\}$ and $|S|$ denotes the cardinality of a set $S$. Furthermore, we write $S^c$ for the complement $[N]\setminus S$ of a set $S$ in $[N]$. $\mathrm{supp}(x):=\{i\in[N]: x_i\neq 0\}$. For a vector $x\in\mathbb{R}^N$ and a set $S\subset [N]$, we denote by $x_S$ the vector coincides with $x$ on the indices in $S$ and is extended to zero outside $S$. For any matrix $A\in\mathbb{R}^{m\times N}$, $\mathrm{ker} A:=\{z\in\mathbb{R}^N: Az=0\}$, $A^T$ is the transpose and $\mathrm{trace}(A)$ is the common trace function. $\langle\cdot,\cdot\rangle$ is the inner product function. $\lfloor\cdot\rfloor$ denotes the floor function.

	\section{$q$-ratio Sparsity and $q$-ratio CMSV}
	
	In this section, we present the definitions of $q$-ratio sparsity and $q$-ratio CMSV, and give their basic properties. A sufficient condition is established for unique sparse recovery via noise free BP. We start with a stable sparsity measure, which is called $q$-ratio sparsity level here. \\

	\begin{definition} (\cite{l1,l2})
		For any non-zero $z\in\mathbb{R}^N$ and non-negative $q\notin\{0,1,\infty\}$, the $q$-ratio sparsity level of $x$ is defined as \begin{align}
		s_{q}(z)=\left(\frac{\lVert z\rVert_1}{\lVert z\rVert_q}\right)^{\frac{q}{q-1}}.
		\end{align}
		The cases of $q\in\{0,1,\infty\}$ are evaluated as limits: 
		\begin{align}
		s_0(z)&=\lim\limits_{q\rightarrow 0} s_q(z)=\lVert z\rVert_0 \\
		s_1(z)&=\lim\limits_{q\rightarrow 1} s_q(z)=\exp(H_1(\pi(z))) \\
		s_\infty(z)&=\lim\limits_{q\rightarrow 0} s_q(z)=\frac{\lVert z\rVert_1}{\lVert z \rVert_\infty}. 
		\end{align}
		Here $\pi(z)\in\mathbb{R}^N$ with entries $\pi_i(z)=|z_i|/\lVert z\rVert_1$ and $H_1$ is the ordinary Shannon entropy. \\
	\end{definition}
	
	This kind of sparsity measure was proposed in \cite{l1,l2}, where estimation and statistical inference via $\alpha$-stable random projection method were studied. Its extension to block sparsity was developed in \cite{zy1}. In fact, this kind of sparsity measure is entropy-based, which counts effective coordinates of $z$ by counting effective states of $\pi(z)$ via entropy. Formally, we have \begin{align}
	s_{q}(z)=\begin{cases}
	\exp(H_q(\pi(z))) &\text{if $z\neq 0$}\\
	0 &\text{if $z=0$},
	\end{cases}
	\end{align}
	where $H_q$ is the R\'{e}nyi entropy of order $q\in[0,\infty]$ \cite{pv,v}. When $q\neq \{0,1,\infty\}$,  the R\'{e}nyi entropy is given by $H_q(\pi(z))=\frac{1}{1-q}\log (\sum_{i=1}^N \pi_i(z)^q)$, and the cases of $q\in\{0,1,\infty\}$ are defined by evaluating limits, with $H_1$ being the
	ordinary Shannon entropy. The sparsity measure $s_{q}(z)$ has the following basic properties (see also \cite{l1,l2}): \\
	\begin{itemize}
		\item Continuity: Unlike the traditional sparsity measure $\ell_0$ norm, the function $s_{q}(\cdot)$ is continuous on $\mathbb{R}^N\setminus \{0\}$ for all $q>0$. Thus, it is stable with respective to small perturbations of the signal.
		\item Range equal to $[0,N]$: For all $z\in\mathbb{R}^N$ and all $q\in [0,\infty]$, we have $$
		0\leq s_{q}(z)\leq N.
		$$
		\item Scale-invariance: For all $c\neq 0$, it holds that $s_{q}(cz)=s_{q}(z)$. This property is in line with the common sense that sparsity should be based on relative (rather than absolute) magnitudes of the entries of the signal.
		\item Non-increasing in $q$: For any $q'\geq q \geq 0$, we have $$
		\frac{\lVert z\rVert_{1}}{\lVert z\rVert_{\infty}}=s_\infty(z)\leq s_{q'}(z)\leq s_{q}(z)\leq s_{0}(z)=\lVert z\rVert_{0},
		$$
		which follows from the non-increasing property of the R\'{e}nyi entropy $H_q$ with respect to $q$. \\
	\end{itemize}
	
	Next, we present a sufficient condition for the exact recovery via noise free BP in terms of $q$-ratio sparsity. First, it is well known that when the true signal $x$ is $k$-sparse, the sufficient and necessary condition for the exact recovery of the noise free BP problem:\begin{align}
	\min\limits_{z\in\mathbb{R}^N}\,\,\lVert z\rVert_1\,\,\,\text{s.t.}\,\,\,Az=Ax \label{nobp}
	\end{align}
	is given by the null space property of order $k$:
	\begin{align*}
	\lVert z_S\rVert_1<\lVert z_{S^c}\rVert_1, \forall z\in\mathrm{ker} A\setminus \{0\}, S\subset [N]\,\text{with}\,|S|\leq k,
	\end{align*}
	see Theorem 4.5 in \cite{fr}. Then, the sufficient condition for exact recovery of $k$-sparse signal via noise free BP (\ref{nobp}) in terms of $q$-ratio sparsity goes as follows. \\
	
	\begin{proposition}
		if $x$ is $k$-sparse and there exists some $1<q\leq\infty$ such that $k$ is strictly less than \begin{align}
		\min\limits_{z\in\mathrm{ker} A\setminus\{0\}}\,\,2^{\frac{q}{1-q}}s_q(z), \label{sufficient}
		\end{align}
		then the unique solution to problem (\ref{nobp}) is the true signal $x$. \\
	\end{proposition}
	
	\noindent
	\emph{Remarks.} Obviously, this result is a direct extension of the Proposition 1 in \cite{tn1}, which states that if the sparsity level $k$ is strictly less than either $\min\limits_{z\in\mathrm{ker} A\setminus\{0\}}\,\,\frac{1}{4}s_2(z)$ or $\min\limits_{z\in\mathrm{ker} A\setminus\{0\}}\,\,\frac{1}{2}s_{\infty}(z)$, then the unique recovery is achieved via (\ref{nobp}). We extends it to a weaker condition, that is $k<\sup\limits_{q\in(1,\infty]} \min\limits_{z\in\mathrm{ker} A\setminus\{0\}}\,\,2^{\frac{q}{1-q}}s_q(z)$. When $q=\infty$, the minimization problem (\ref{sufficient}) can be solved by solving $N$ linear programs with a polynomial time, see the algorithm (\ref{linfty}). However, in the cases of $1<q<\infty$, it's very difficult to solve exactly. In Section V, we adopt a convex-concave procedure algorithm to solve it approximately. \\
	
	Now we are ready to present the definition of $q$-ratio constrained minimal singular value, which is developed based on $q$-ratio sparsity level.\\
	
	\begin{definition}
		For any real number $s\in[1,N]$, $1<q\leq \infty$ and matrix $A\in\mathbb{R}^{m\times N}$, the $q$-ratio constrained minimal singular value (CMSV) of $A$ is defined as \begin{align}
		\rho_{q,s}(A)=\min\limits_{z\neq 0,s_q(z)\leq s}\,\,\frac{\lVert Az\rVert_2}{\lVert z\rVert_q}.
		\end{align}
	\end{definition}
	
	\noindent\\
	\emph{Remarks.} When $q=2$ and $q=\infty$, $\rho_{q,s}(A)$ reduces to $\rho_s(A)$ given in \cite{tn1} and $w_2(A,s)$ given in \cite{tn3}, respectively. For measurement matrices $A$ with columns of unit norm, it is obvious that $\rho_{q,s}(A)\leq 1$ for any $q>1$ since $\lVert Ae_i\rVert_2=1$, $\lVert e_i\rVert_q=1$ and $s_q(e_i)=1$, where $e_i$ is the $i$-th canonical basis for $\mathbb{R}^N$. Moreover, when $q$ and $A$ are fixed, as a function with respective to $s$, $\rho_{q,s}(A)$ is non-increasing. For any $\alpha\in\mathbb{R}$, we have $\rho_{q,s}(\alpha A)=|\alpha|\rho_{q,s}(A)$. This fact together with Theorem 1 in Section III imply that increasing the sensing energy without changing the sensing matrix structure proportionally reduces the $\ell_q$ norm of reconstruction errors when the true signal is exactly sparse. In fact, other extensions of this definition can be done, for instance we can define, for any $1<q\leq\infty$, \begin{align*}
	\rho_{\lozenge,q}(A,s)=\min\limits_{z\neq 0,s_q(z)\leq s}\,\,\frac{\lVert Az\rVert_{\lozenge}}{\lVert z\rVert_q},
	\end{align*} 
	where $\lVert \cdot\rVert_{\lozenge}$ denotes a general norm. Then the measure $\omega_{\lozenge}(A,s)$ defined in \cite{tn3} is exactly $\rho_{\lozenge,\infty}(A,s)$. Thus the corresponding results there can be generalized in terms of this newly defined measure. But we do not pursue the extensions in this paper. Basically, the recovery condition ($\rho_{2,s}(A)>0$ with some proper $s$) discussed later to achieve the $\ell_2$ norm error bounds is equivalent to the robust width property investigated in \cite{cm,z,zy2}. The recovery condition in terms of $q$-ratio CMSV is quiet similar to the restricted eigenvalue condition \cite{brt,rwy} or more general restricted strong convexity condition \cite{nywr}. The difference is that here we use a restricted set in terms of $q$-ratio sparsity, which is more intuitive and makes the proof procedure more concise. Not least it is computable! Comparing to the RIP, all these conditions do not require upper bounds on the restricted eigenvalues.  \\
	
	As for different $q$, we have the following important inequality, which will play a crucial role in analysing the probabilistic behavior of $\rho_{q,s}(A)$ via the existing results for $\rho_{2,s}(A)$ established in \cite{tn1}. \\
	
	\begin{proposition}
		If $1<q_2\leq q_1\leq\infty$, then for any real number $1\leq s\leq N^{\frac{q_1(q_2-1)}{q_2(q_1-1)}}$, we have \begin{align}
		\rho_{q_1,s}(A)\geq \rho_{q_2,s^{\frac{q_2(q_1-1)}{q_1(q_2-1)}}}(A)\geq  s^{-\frac{q_2(q_1-1)}{q_1(q_2-1)}} \rho_{q_1, s^{\frac{q_2(q_1-1)}{q_1(q_2-1)}}}(A). \label{rhoineq}
		\end{align}
	\end{proposition}
	
	\noindent\\
	\emph{Remarks.} Let $q_1=\infty$ and $q_2=2$, we have $\rho_{\infty,s}(A)\geq \rho_{2,s^2}(A)\geq \frac{1}{s^2}\rho_{\infty,s^2}$. The left hand side is exactly the right hand side inequality of Proposition 4 in \cite{tn3}, i.e., $w_2(A,s)\geq \rho_{s^2}(A)$. But as $\frac{q_2(q_1-1)}{q_1(q_2-1)}=1+\frac{q_1-q_2}{q_1(q_2-1)}\geq1$ as $q_1\geq q_2>1$, so $\rho_{q_2,s^{\frac{q_2(q_1-1)}{q_1(q_2-1)}}}(A)\leq \rho_{q_2,s}(A)$. Similarly, according to the right hand side of the inequality, we have for any $t\in[1,N]$ $\rho_{q_2,t}(A)\geq \frac{1}{t}\rho_{q_1,t}(A)$. But obviously $\frac{1}{t}\rho_{q_1,t}(A)\leq \rho_{q_1,t}(A)$. Therefore, we can not obtain the monotonicity with respective to $q$ of $\rho_{q,s}(A)$ when $s$ and $A$ are fixed. However, when $s=N$, then since for any $z\in\mathbb{R}^N$, $s_q(z)\leq N$, it holds trivially that $\rho_{q,N}(A)$ is increasing with respect to $q$ by using the decreasing property of $\ell_q$ norm.
	
	\section{Recovery results}
	
	In this section, we derive performance bounds on both $\ell_q$ norm and $\ell_1$ norm of the reconstruction errors for several convex sparse recovery algorithms in terms of the $q$-ratio CMSV of the measurement matrix. Let $y=Ax+w\in\mathbb{R}^m$ where $x\in\mathbb{R}^N$ is the true sparse or compressible signal, $A\in\mathbb{R}^{m\times N}$ is the measurement matrix and $w\in\mathbb{R}^m$ is the noise vector. We focus on three most renowned sparse recovery algorithms based on convex relaxation: the BP, the DS and the Lasso estimator. \\
	
	BP: $\min\limits_{z\in\mathbb{R}^N}\,\,\lVert z\rVert_1\,\,\,\text{s.t.}\,\,\,\lVert y-Az\rVert_2\leq \varepsilon$. \\
	
	DS: $\min\limits_{z\in\mathbb{R}^N}\,\,\lVert z\rVert_1\,\,\,\text{s.t.}\,\,\,\lVert A^{T}(y-Az)\rVert_\infty\leq \lambda_N \sigma$.\\
	
	Lasso: $\min\limits_{z\in\mathbb{R}^N}\frac{1}{2}\lVert y-Az\rVert_2^2+\lambda_N\sigma\lVert z\rVert_1$.\\
	
	Here $\varepsilon$, $\lambda_N$ and  $\sigma$ are parameters used in the conditions to control the noise levels. We first present the following main recovery results for the case that the true signal $x$ is exactly sparse. \\
	
	\begin{theorem}
		Suppose $x$ is $k$-sparse. For any $1<q\leq \infty$, we have  \\
		1) If $\lVert w\rVert_2\leq \varepsilon$, then the solution $\hat{x}$ to the BP obeys \begin{align}
		\lVert\hat{x}-x\rVert_{q}&\leq \frac{2\varepsilon}{\rho_{q,2^{\frac{q}{q-1}}k}(A)},  \label{noisebp} \\
		\lVert\hat{x}-x\rVert_{1}&\leq \frac{4k^{1-1/q}\varepsilon}{\rho_{q,2^{\frac{q}{q-1}}k}(A)}.
		\end{align}
		2) If the noise $w$ in the DS satisfies $\lVert A^T w\rVert_{\infty}\leq \lambda_N \sigma$, then the solution $\hat{x}$ to the DS obeys \begin{align}
		\lVert\hat{x}-x\rVert_{q}\leq \frac{4k^{1-1/q}}{\rho_{q,2^{\frac{q}{q-1}}k}^2(A)}\lambda_N\sigma, \\
		\lVert\hat{x}-x\rVert_{1}\leq \frac{8k^{2-2/q}}{\rho_{q,2^{\frac{q}{q-1}}k}^2(A)}\lambda_N\sigma.
		\end{align}
		3) If the noise $w$ in the Lasso satisfies $\lVert A^T w\rVert_{\infty}\leq \kappa \lambda_N\sigma$ for some $\kappa\in(0,1)$, then the solution $\hat{x}$ to the Lasso obeys \begin{align}
		\lVert\hat{x}-x\rVert_{q}&\leq \frac{1+\kappa}{1-\kappa}\cdot\frac{2k^{1-1/q}}{\rho_{q,(\frac{2}{1-\kappa})^{\frac{q}{q-1}}k}^2(A)}\lambda_N\sigma, \\
		\lVert\hat{x}-x\rVert_{1}&\leq \frac{1+\kappa}{(1-\kappa)^2}\cdot\frac{4k^{2-2/q}}{\rho_{q,(\frac{2}{1-\kappa})^{\frac{q}{q-1}}k}^2(A)}\lambda_N\sigma.\label{sparselassol1}  
		\end{align}
	\end{theorem}
	
	\noindent
	\emph{Remarks.} When the noise vector $w\sim N(0,\sigma^2 I_m)$, the conditions on noise for the DS and Lasso hold with high probability if $\lambda_N$ (the parameter related to the signal dimensional $N$) is properly chosen. As a by product of (\ref{noisebp}), we have if $\rho_{q,2^{\frac{q}{q-1}}k}(A)>0$, then the noise free BP (\ref{nobp}) can uniquely recover any $k$-sparse  signal by letting $\varepsilon=0$. We established both the error $\ell_q$ norm and $\ell_1$ norm bounds. Our results for the error $\ell_q$ norm bounds generalize from the existing results in \cite{tn1} ($q=2$) and \cite{tn3} ($q=\infty$) to any $1<q\leq \infty$. The error $\ell_q$ norm bounds depend on the $q$-ratio CMSV of the measurement matrix $A$, which is bounded away from zero for subgaussian random matrices and can be computed approximately by using some specific algorithms. The details will be discussed in the later sections.\\
	
	Next, we extend Theorem 1 to the case that the true signal is allowed to be not exactly sparse, but is compressible, i.e., it can be well approximately by an exactly sparse signal.\\
	
	\begin{theorem}
		Let the $\ell_1$-error of best $k$-term approximation of $x$ be $\sigma_k(x)_1=\inf \{\lVert x-z\rVert_1, z\in\mathbb{R}^N \text{is $k$-sparse}\}$, which is a function that measures how close $x$ is to being $k$-sparse. For any $1<q\leq \infty$, we have  \\
		1) If $\lVert w\rVert_2\leq \varepsilon$, then the solution $\hat{x}$ to the BP obeys \begin{align}
		\lVert\hat{x}-x\rVert_{q}&\leq \frac{2\varepsilon}{\rho_{q,4^{\frac{q}{q-1}}k}(A)}+k^{1/q-1}\sigma_k(x)_1 \label{robust1}, \\
		\lVert\hat{x}-x\rVert_{1}&\leq \frac{4k^{1-1/q}\varepsilon}{\rho_{q,4^{\frac{q}{q-1}}k}(A)}+4\sigma_k(x)_1. \label{robust1l1}
		\end{align}
		2) If the noise $w$ in the DS satisfies $\lVert A^T w\rVert_{\infty}\leq \lambda_N \sigma$, then the solution $\hat{x}$ to the DS obeys \begin{align}
		\lVert\hat{x}-x\rVert_{q}&\leq \frac{8k^{1-1/q}}{\rho_{q,4^{\frac{q}{q-1}}k}^2(A)}\lambda_N\sigma+k^{1/q-1}\sigma_k(x)_1 \label{robust2}, \\
		\lVert\hat{x}-x\rVert_{1}&\leq \frac{16k^{2-2/q}}{\rho_{q,4^{\frac{q}{q-1}}k}^2(A)}\lambda_N\sigma+4\sigma_k(x)_1. \label{robust2l1}
		\end{align}
		3) If the noise $w$ in the Lasso satisfies $\lVert A^T w\rVert_{\infty}\leq \kappa \lambda_N\sigma$ for some $\kappa\in(0,1)$, then the solution $\hat{x}$ to the Lasso obeys \begin{align}
		\lVert\hat{x}-x\rVert_{q}&\leq \frac{1+\kappa}{1-\kappa}\cdot\frac{4k^{1-1/q}}{\rho_{q,(\frac{4}{1-\kappa})^{\frac{q}{q-1}}k}^2(A)}\lambda_N\sigma+k^{1/q-1}\sigma_k(x)_1 \label{robust3}, \\
		\lVert\hat{x}-x\rVert_{1}&\leq \frac{1+\kappa}{(1-\kappa)^2}\cdot\frac{8k^{2-2/q}}{\rho_{q,(\frac{4}{1-\kappa})^{\frac{q}{q-1}}k}^2(A)}\lambda_N\sigma+\frac{4}{1-\kappa}\sigma_k(x)_1. \label{robust3l1} 
		\end{align}
	\end{theorem}
	
	\noindent
	\emph{Remarks.} As we can see, all the error bounds consist of two components, one is caused by the measurement error, while the other one is caused by the sparsity defect.  And according to the proof procedure presented later, we can sharpen the error bounds to be the maximum of these two components instead of their summation. Comparing to the exactly sparse case, we need slightly stronger conditions to achieve the valid error bounds. Concisely, we require $\rho_{q,4^{\frac{q}{q-1}}k}(A)>0$, $\rho_{q,4^{\frac{q}{q-1}}k}(A)>0$ and $\rho_{q,(\frac{4}{1-\kappa})^{\frac{q}{q-1}}k}(A)>0$ for the BP, DS and Lasso in the compressible case, while the conditions are $\rho_{q,2^{\frac{q}{q-1}}k}(A)>0$, $\rho_{q,2^{\frac{q}{q-1}}k}(A)>0$ and $\rho_{q,(\frac{2}{1-\kappa})^{\frac{q}{q-1}}k}(A)>0$ in the exactly sparse case, respectively.\\

	\section{Random matrices}
	
	In this section, we study the property of $q$-ratio CMSVs for the subgaussian random matrices. A random vector $X\in\mathbb{R}^N$ is called isotropic and subgaussian with constant $L$ if it holds for all $u\in\mathbb{R}^N$ that $E|\langle X,u\rangle|^2=\lVert u\rVert_2^2$ and $P(|\langle X, u\rangle|\geq t)\leq 2\exp(-\frac{t^2}{L\lVert u\rVert_2})$. Then as shown in Theorem 2 of \cite{tn1}, we have the following lemma. \\
	
	\begin{lemma} (\cite{tn1})
		Suppose the rows of the scaled measurement matrix $\sqrt{m}A$ to be i.i.d isotropic and subgaussian  random vectors with constant $L$. Then there exists constants $c_1$ and $c_2$ such that for any $\eta>0$ and $m\geq 1$ satisfying $$
		m\geq c_1\frac{L^2s\log N}{\eta^2}
		$$
		we have $$
		E|1-\rho_{2,s}(A)|\leq \eta
		$$
		and $$
		P(1-\eta\leq\rho_{2,s}(A)\leq 1+\eta)\geq 1-\exp(-c_2\eta^2\frac{m}{L^4}).
		$$ \\
	\end{lemma}
	
	Then as a direct consequence of Proposition 2 (i.e., if $1<q<2$, $\rho_{q,s}(A)\geq s^{-1}\rho_{2,s}(A)$. While if $2\leq q\leq \infty$, $\rho_{q,s}(A)\geq\rho_{2,s^{\frac{2(q-1)}{q}}}(A)$.) and Lemma 1, we have the following probabilistic statements about $\rho_{q,s}(A)$. \\
	
	\begin{theorem}
		Under the assumptions and notations of Lemma 1, it holds that \\
		
		\noindent
		1) When $1<q< 2$, there exist constants $c_1$ and $c_2$ such that for any $\eta>0$ and $m\geq 1$ satisfying $$
		m\geq c_1\frac{L^2 s\log N}{\eta^2}
		$$
		we have \begin{align}
		E[\rho_{q,s}(A)]&\geq s^{-1}(1-\eta), \\
		P\{\rho_{q,s}(A)&\geq s^{-1}(1-\eta)\}\geq 1-\exp(-c_2\eta^2 \frac{m}{L^4}).  \\ \nonumber 
		\end{align}
		
		\noindent
		2) When $2\leq q\leq \infty$, there exist constants $c_1$ and $c_2$ such that for any $\eta>0$ and $m\geq 1$ satisfying $$
		m\geq c_1\frac{L^2 s^{\frac{2(q-1)}{q}}\log N}{\eta^2}
		$$
		we have \begin{align}
		E[\rho_{q,s}(A)]&\geq 1-\eta, \\
		P\{\rho_{q,s}(A)&\geq 1-\eta\}\geq 1-\exp(-c_2\eta^2 \frac{m}{L^4}).
		\end{align}\\
		
	\end{theorem} 
	
	\noindent
	\emph{Remarks.} Theorem 3 shows that at least for subgaussian random matrices, the $q$-ratio CMSV is bounded away from zero as long as the number of measurements is large enough. Random measurement matrices with i.i.d isotropic subgaussian random vector rows include the Gaussian and Bernoulli ensembles. The order of required number of measurements $m$ are close to the optimal order for establishing the $\ell_q$ norm error bound, see \cite{dlr}. Besides, \cite{zc} shows that the $\rho_{2,s}(A)$ of a class of structured random matrices including the Fourier random matrices and Hadamard random matrices is bounded from zero with high probability as long as the number of measurements is reasonably large. Then by adopting Proposition 2 again, this conclusion still holds for the $\rho_{q,s}(A)$ with $1<q\leq \infty$. \\
	
\begin{figure}[htbp]
		\centering
		\includegraphics[width=0.9\textwidth,height=0.45\textheight]{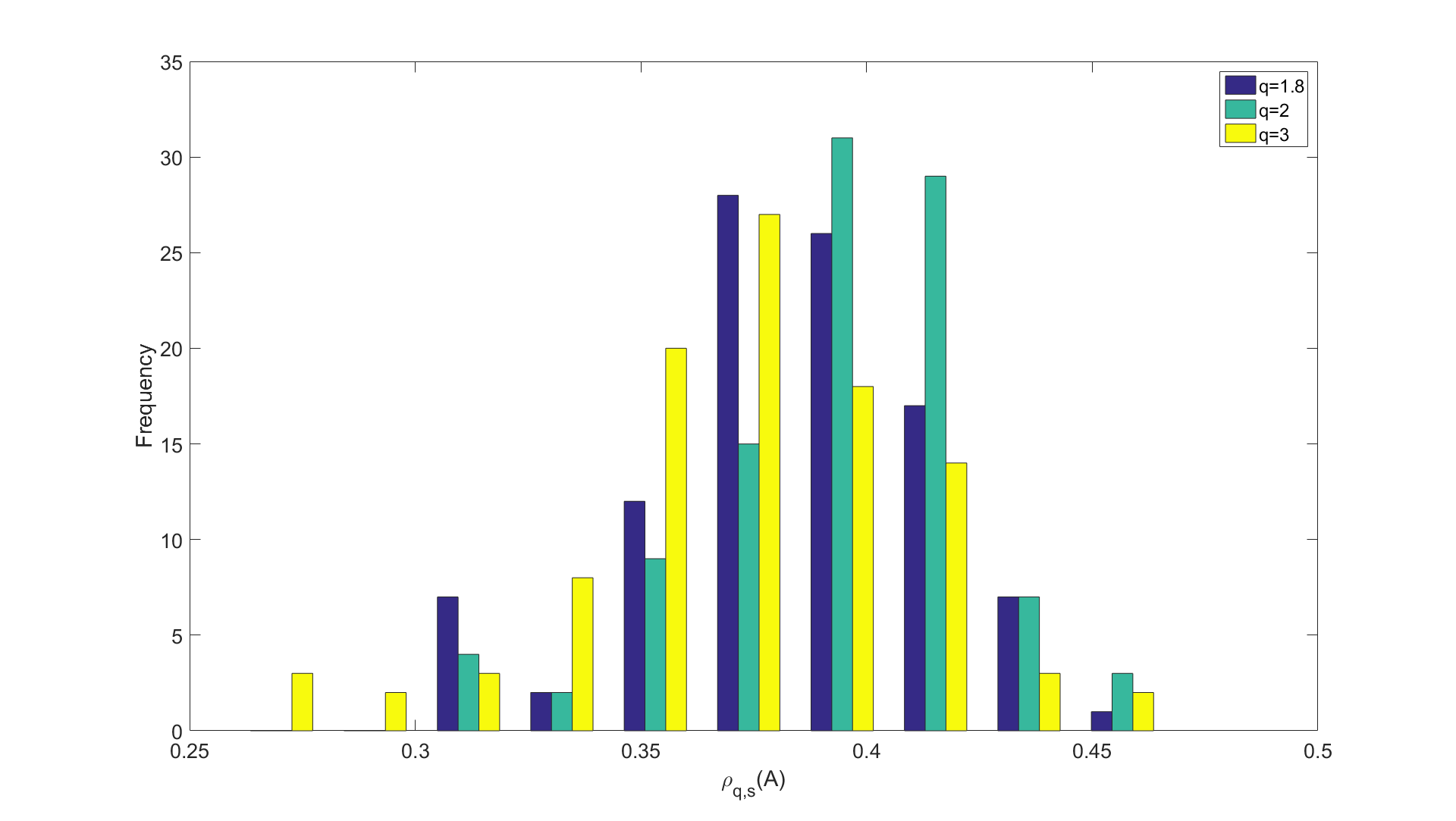}
		\caption{Histograms of the $q$-ratio CMSVs for Gaussian random matrices of size $40\times 60$ with $s=4$ and $q=1.8,2,3$.}\label{fig:1}
	\end{figure}

	In Fig.\,\ref{fig:1}, we plot the histograms of $\rho_{q,s}(A)$ using the computing algorithm (\ref{IP}) for Gaussian random matrices $A\in\mathbb{R}^{40\times 60}$ normalized by $\frac{1}{\sqrt{40}}$. We set $s=4$ but with three different kind of $q$, i.e., $q=1.8$, $q=2$ and $q=3$. We obtain each histogram from 100 Gaussian random matrices. It can be observed that as is expected that the $q$-ratio CMSVs are all bounded away from zero both in expectation and with high probability in this setting. 
	
	\section{Numerical experiments}
	In this section, we first describe a convex-concave procedure used to compute the maximal sparse level $k$ such that the sufficient condition (\ref{sufficient}) is fulfilled, and then introduce the computation of the $q$-ratio CMSV and compare the $q$-ratio CMSV based bound and RIC based bound on the BP.
	
	\subsection{Verifying Sufficient Conditions}
	In order to use the $q$-ratio sparsity level to verify the sufficient condition (\ref{sufficient}), for each $1<q\leq \infty$, we need to solve the optimization problem: \begin{align}
	\min\limits_{z\in\mathrm{ker} A\setminus\{0\}}\,\,2^{\frac{q}{1-q}}\left(\frac{\lVert z\rVert_1}{\lVert z\rVert_q}\right)^{\frac{q}{q-1}}.
	\end{align}
	Regardless of the constant, it is essentially equivalent to solve the problem: \begin{align}
	\max\limits_{z\in\mathbb{R}^N}\,\lVert z\rVert_q\,\,\,\text{s.t. $Az=0$ and $\lVert z\rVert_1\leq 1$}. \label{maxq}
	\end{align}
	Unfortunately, this maximizing $\ell_q$ norm over a polyhedron problem is non-convex. For $q=2$, \cite{tn1} proposed to use a semidefinite relaxation to obtain an upper bound: \begin{align}
	(L_2): &\max\limits_{Z\in\mathbb{R}^{N\times N}:Z\succeq 0}\,\mathrm{trace}(Z) \nonumber   \\ 
	&\text{s.t.}\,\,\mathrm{trace}(AZA^T)=0, \lVert Z\rVert_1\leq 1, \label{l2}
	\end{align}
	where $Z=zz^T$ and $\lVert Z\rVert_1$ is the entry-wise $\ell_1$ norm of $Z$. For $q=\infty$, it can be solved by solving $N$ linear programs (see \cite{tn1,tn3}): \begin{align}
	(L_\infty): \max\limits_{1\leq i\leq N}\{\max\limits_{z\in\mathbb{R}^N} z_i,\,\, \text{s.t.\,\,$Az=0$ and $\lVert z\rVert_1\leq 1$}\}. \label{linfty}
	\end{align}
	Here we adopt the convex-concave procedure (CCP) (see \cite{lb} for details) to solve the problem (\ref{maxq}) for any $1<q<\infty$. The basic CCP algorithm goes as follows: \\
	
	\fbox{%
		\parbox{0.95\textwidth}{%
			{\emph{Algorithm:}} CCP to solve (\ref{maxq}).\\ 
			Given an initial point $z_0$. Let $k=0$.\\
			Repeat\\
			1. Convexify. Linearize $\lVert z\rVert_q$ with the approximation \begin{align*}\lVert z_k\rVert_q+\nabla (\lVert z\rVert_q)_{z=z_k}^T(z-z_k)=\lVert z_k\rVert_q+[\lVert z_k\rVert_q^{1-q}|z_k|^{q-1}\mathrm{sign}(z_k)]^T(z-z_k).
			\end{align*}
			2. Solve. Set the value of $z_{k+1}$ to be a solution of \begin{align}
			&\max\limits_{z}\,\lVert z_k\rVert_q+(\lVert z_k\rVert_q^{1-q}|z_k|^{q-1}\mathrm{sign}(z_k))^T(z-z_k)  \nonumber \\
			&\quad\text{s.t.}\,\,Az=0, \lVert z\rVert_1\leq 1. \label{ccp}
			\end{align}
			3. Update iteration: $k=k+1$.\\
			Until stopping criterion is satisfied. 
		}%
	} \bigskip
	
	We first compare the $L_2$ and $L_{\infty}$ algorithms for verifying the sufficient condition with our developed CCP algorithm. We present the results of CCP algorithms for $q=1.8,2,3$ and 20 here. All the convex problems including (\ref{l2}), (\ref{linfty}) and (\ref{ccp}) are solved by CVX toolbox in Matlab \cite{gby}. The initial point $z_0$ used in CCP is taken to be the solution of $L_\infty$ (\ref{linfty}). If we denote the optimal objective values obtained to solve  (\ref{l2}), (\ref{linfty}) and (\ref{maxq}) via CCP algorithm with some $q$ by $L_{2}^{optival}$, $L_{\infty}^{optival}$ and $CCP_{q}^{optival}$, then the maximal sparsity levels to achieve unique recovery for noise free BP are calculated as $\lfloor 1/(4L_{2}^{optival})\rfloor$, $\lfloor 1/(2L_{\infty}^{optival})\rfloor$ and $\lfloor 2^{\frac{q}{1-q}}(1/CCP_{q}^{optival})^{\frac{q}{q-1}}\rfloor$, respectively. In TABLE \ref{table:1}, we present the corresponding maximal sparsity levels calculated via different algorithms for a small size Bernoulli matrix with fixed $N=40$ while varying the number of measurements $m$. As is shown, the convex relaxation algorithm $L_2$ actually give a lower bound for the solution of the case $q=2$. In TABLE \ref{table:2}, we compare the results computed by $L_{\infty}$ and CCP for larger Gaussian random matrix with $N=256$, also varying the number of measurements $m$. From both tables, it is observed that the maximal sparsity levels to achieve unique recovery for noise free BP computed by the algorithms $L_2$ and $L_\infty$ are quite conservative. But it is much more acceptable and closer to the theoretical well-known optimal recovery condition bound $m\geq 2k\ln(N/k)$ for the Gaussian measurement matrix via our proposed CCP algorithm with some proper $q$, for instance $q=2$. According to Proposition 1, to obtain unique recovery for noise free BP, the sparsity level of the unknown true signal is merely required to less than or equal to the maximal sparsity level calculated for some $q$. \\
	
	\begin{table}[h]
		\caption{Comparison of the maximal sparsity levels calculated via different algorithms for a Bernoulli matrix with $N=40$.} 
		\centering 
		\begin{tabular}{c|c|c|c|c|c|c} 
			\hline\hline 
			$m$ & $L_2$ & $L_\infty$ & $\mathrm{CCP}_{1.8}$ & $\mathrm{CCP}_{2}$ & $\mathrm{CCP}_{3}$ & $\mathrm{CCP}_{20}$ \\ 
			\hline 
			20 & 1 & 1 & 1 & 1 & 2 & 2\\ 
			24 & 1 & 2 & 2 & 2 & 3 & 2\\
			28 & 2 & 2 &  2 & 3 & 3 & 2\\
			32 & 2 & 3 & 3 & 3 & 4 & 3\\
			\hline \hline 
		\end{tabular}
		\label{table:1} 
	\end{table}

	\begin{table}[h]
		\caption{Comparison of the maximal sparsity levels calculated via different algorithms for a Gaussian matrix with $N=256$.} 
		\centering 
		\begin{tabular}{c|c|c|c|c|c} 
			\hline\hline 
			$m$  & $L_\infty$ & $\mathrm{CCP}_{1.8}$ & $\mathrm{CCP}_{2}$ & $\mathrm{CCP}_{3}$& $\mathrm{CCP}_{20}$ \\ 
			\hline 
			25 &  1 & 1 & 1 & 1 &1\\ 
			51 &  2 & 2 & 3 & 3 &2\\
			76 &  2 & 4 & 4 & 4 &2\\
			102 &  3 & 7 & 7 & 6 &3\\
			128 &  4 & 10 & 10 & 9 &4\\
			153 &  5 & 13 & 14 & 12 &6\\
			179 &  7 & 17 & 18 & 16 &7\\
			204 &  9 & 20 & 23 & 23 &10\\
			230 &  12 & 27 & 31 & 32 &13\\
			\hline \hline 
		\end{tabular}
		\label{table:2} 
	\end{table}

	\subsection{Computing $q$-ratio CMSVs}
	For each $q$, the computation of the $q$-ratio CMSV is equivalent to \begin{align}
	\min\limits_{z\in\mathbb{R}^N}\,\lVert Az\rVert_2\,\,\,\text{s.t.}\,\,\,\lVert z\rVert_1\leq s^{\frac{q-1}{q}}, \lVert z\rVert_q=1. \label{computing}
	\end{align}
	
	The above optimization problem is not convex because of the $\ell_q$ constraint $\lVert z\rVert_q=1$. Here we use an interior point (IP) algorithm to directly compute an approximate numerical solution of (\ref{computing}). However IP approach requires that the objective and constraint function to possess continuous second order derivatives, which is not fulfilled by the constraint $\lVert z\rVert_1-s^{\frac{q-1}{q}}\leq 0$. The problem can be addressed by defining $z=z^{+}-z^{-}$ with $z^{+}=\max(z,0)$ and $z^{-}=\max(-z,0)$. This leads to the following augmented optimization problem: \begin{align}
	\min\limits_{z^{+},z^{-}\in\mathbb{R}^N}&\,(z^{+}-z^{-})^T A^T A(z^{+}-z^{-}) \nonumber \\
	&\text{s.t.}\,\,\,\sum\limits_{i}z_i^{+}+\sum\limits_{i}z_i^{-}-s^{\frac{q-1}{q}}\leq 0, \nonumber \\
	&\lVert z^{+}-z^{-}\rVert_q^q=1, \nonumber\\
	&z^{+}\geq 0, z^{-}\geq 0. \label{IP} 
	\end{align} 
	
\begin{figure}[htbp]
		\centering
		\includegraphics[width=0.9\textwidth,height=0.45\textheight]{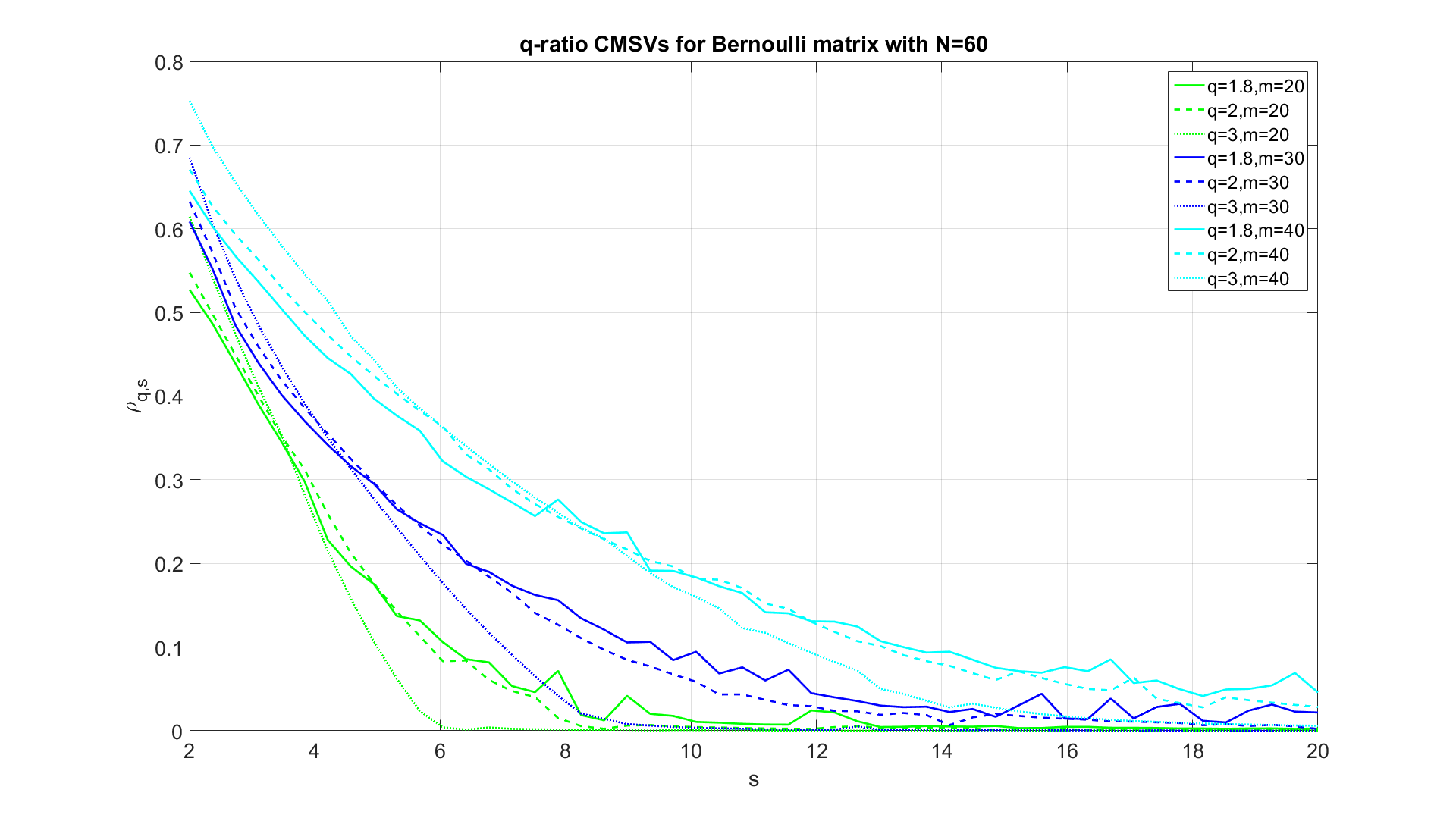}
		\caption{$q$-ratio CMSV $\rho_{q,s}$ for Bernoulli random matrices of size $60$ as a function of $s$ with $q=1.8,2,3$ and $m=20,30,40$.}\label{fig:2}
	\end{figure}

	The IP algorithm is implemented using the Matlab function fmincon. Due to the existences of local minima, we run the IP 30 times and select the minimal function value for all the trials. In Fig. \ref{fig:2}, we compare the $q$-ratio CMSVs as a function of $s$ approximated by the IP for Bernoulli random matrices. We set $N=60$ but with three different $m=20,30,40$ and three different $q=1.8,2,3$. A Bernoulli random matrix with dimensionality $40\times 60$ is first simulated. Then for $m=20,30,40$, the corresponding matrix is obtained by taking the first $m$ rows of that full Bernoulli random matrix. And the columns of all the used matrix are normalized to have unit norms, which guarantees that $\rho_{q,s}(A)\leq 1$ for any $q>1$. In general, as is shown that the $q$-ratio CMSVs decrease as $s$ increases for all the nine cases. For fixed $s$, the $q$-ratio CMSVs increases as $m$ increases for all the $q$. The influence of $q$ on the $q$-ratio CMSVs is relatively small and apparently not monotonous.
	
\begin{figure}[htbp]
		\centering
		\includegraphics[width=0.9\textwidth,height=0.45\textheight]{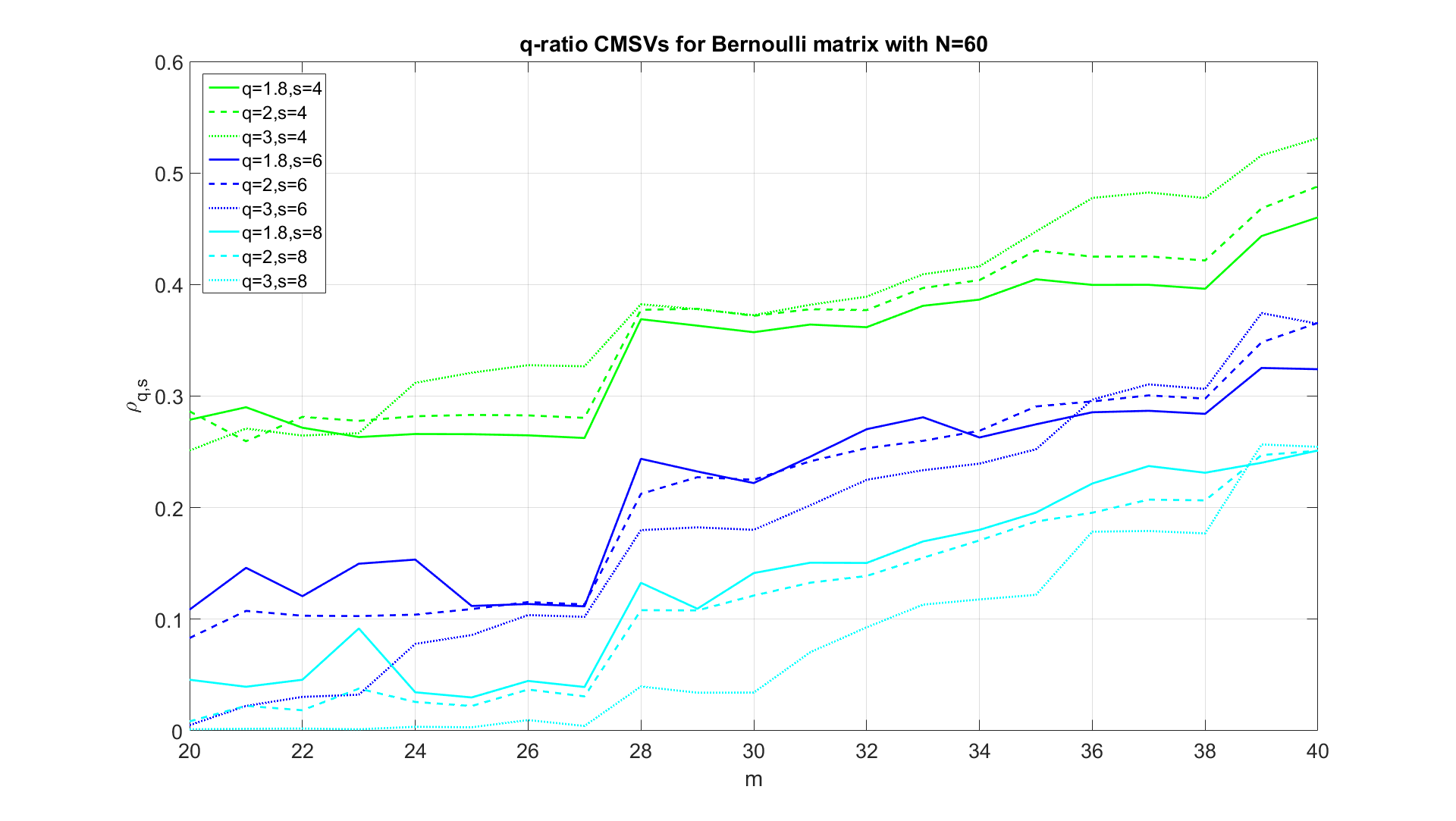}
		\caption{$q$-ratio CMSV $\rho_{q,s}$ for Bernoulli random matrices of size $60$ as a function of $m$ with $q=1.8,2,3$ and $s=4,6,8$.}\label{fig:3}
	\end{figure}

\begin{figure}[htbp]
		\centering
		\includegraphics[width=0.9\textwidth,height=0.45\textheight]{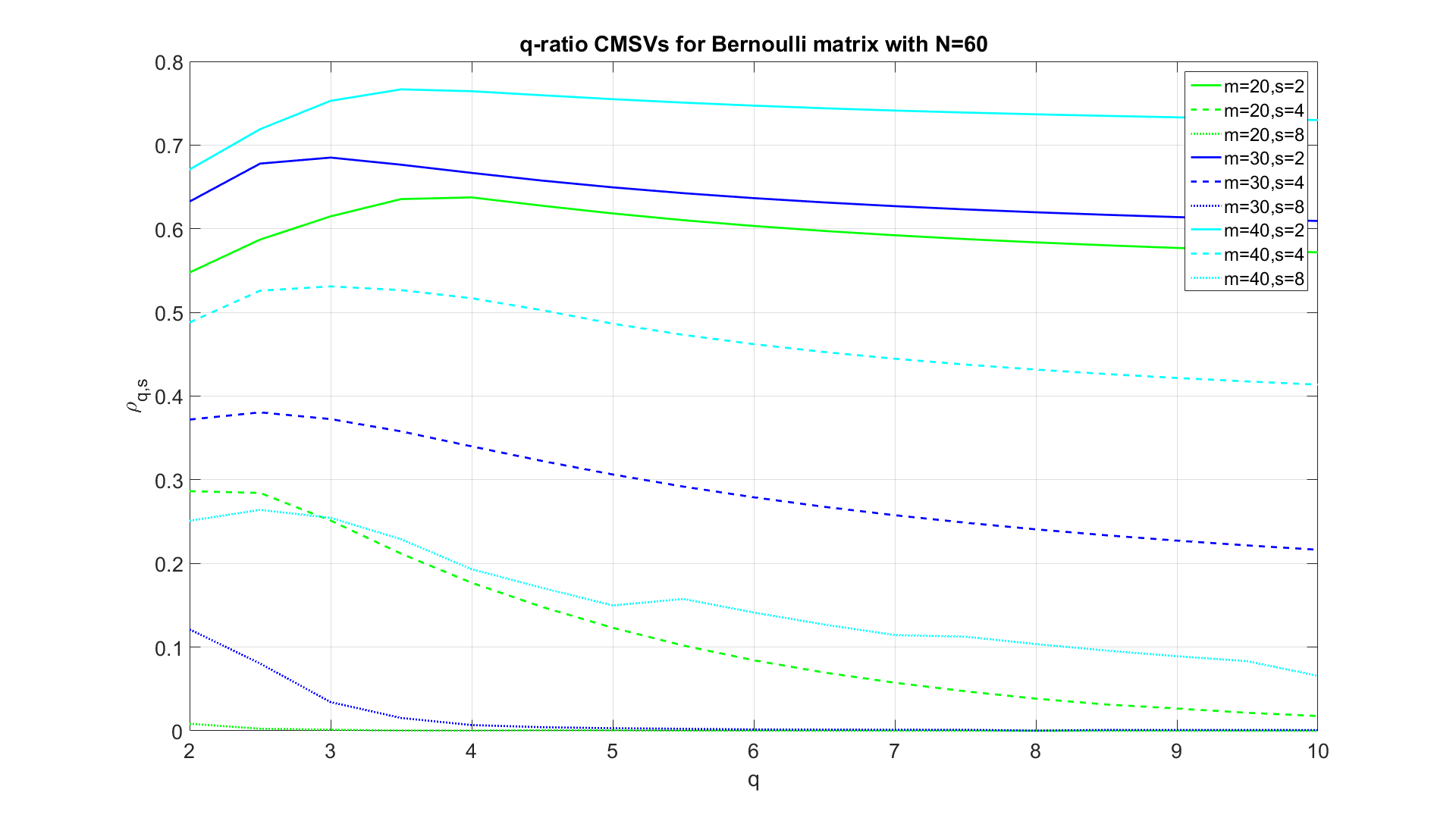}
		\caption{$q$-ratio CMSV $\rho_{q,s}$ for Bernoulli random matrices of size $60$ as a function of $q$ with $m=20,30,40$ and $s=2,4,8$.}\label{fig:4}
	\end{figure}

	In Fig. \ref{fig:3}, we plot the $q$-ratio CMSVs as a function of $m$ with varying $q=1.8,2,3$ and $s=4,6,8$. For each $q$ and $s$, we computing the $q$-ratio CMSVs with $m$ increasing from 20 to 40. For each $m$, the construction of the corresponding matrix follows the same procedure given previously. Under the same settings, the $q$-ratio CMSVs as a function of $q$ with varying $m=20,30,40$ and $s=2,4,8$ are presented in Fig. \ref{fig:4}. Similar behaviors as Fig. \ref{fig:2} are observed from these two figures. 
	
	\subsection{Bounds Comparison}
	Finally, we compare the $q$-ratio CMSV based bound and the RIC based bound on the BP for different configurations of $m$ and $k$. It's known that if the $2k$ order RIC of the measurement matrix $A$ satisfies that $\delta_{2k}(A)<\sqrt{2}-1$, then for any solution $\hat{x}$ of the noisy BP approximates the true $k$-sparse signal $x$ with errors \begin{align}
	\lVert x-\hat{x}\rVert_{q}\leq Ck^{1/q-1/2}\varepsilon,
	\end{align}
	where $C=\frac{4\sqrt{1+\delta_{2k}(A)}}{1-(1+\sqrt{2})\delta_{2k}(A)}$ with any $1\leq q\leq 2$.
	
	Without loss of generality, we set $\varepsilon=1$. The RIC is approximated using Monte Carlo simulations. Specifically, to compute $\delta_{2k}(A)$, we randomly take 1000 submatrices of $A\in\mathbb{R}^{m\times N}$ of size $m\times 2k$, and approximate $\delta_{2k}(A)$ using the maximum of $\max(\sigma_1^2-1,1-\sigma_{2k}^2)$ among all sampled submatrices. Here $\sigma_1$ and $\sigma_{2k}$ are the corresponding maximal and minimal singular values of the sampled submatrix. As it is obvious that the approximated RIC is always smaller than or equal to the exact RIC, the error bounds based on the exact RIC are always worse than those based on the approximated RIC. Therefore, if our $q$-ratio CMSV based bound is better than the approximated RIC based bound, it is even better than the exact RIC based one.
	
\begin{figure}[htbp]
		\centering
		\includegraphics[width=0.9\textwidth,height=0.45\textheight]{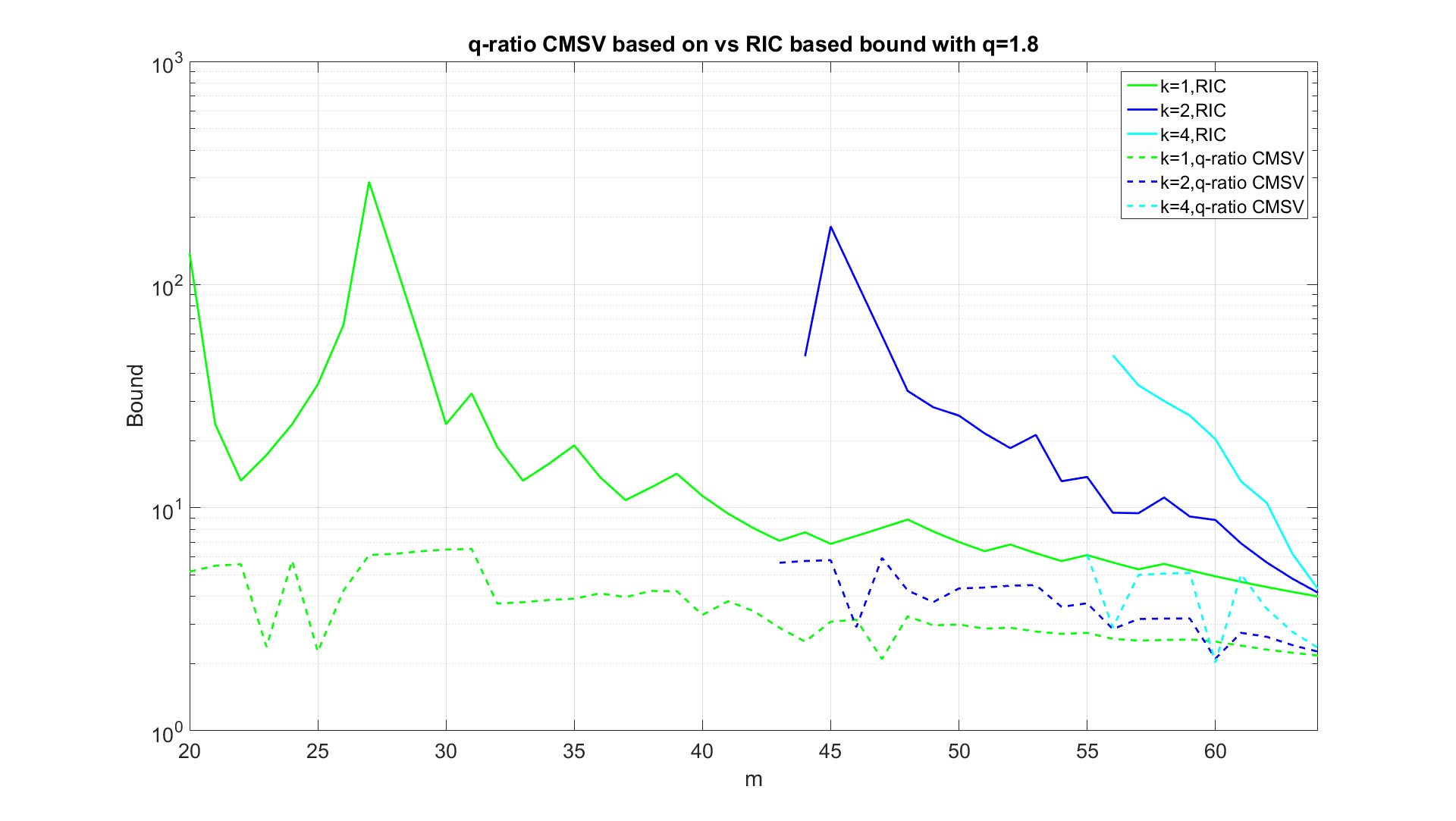}
		\caption{The $q$-ratio CMSV $\rho_{q,s}$ based bound vs the RIC based bound for Hadamard submatrices with $N=64$, $k=1,2,4$ and $q=1.8$.}\label{fig:5}
	\end{figure}

	We approximate the $q$-ratio CMSV and the RIC for column normalized submatrices of a row-randomly-permuted Hadamard matrix with $N=64$, $k=1,2,4$, $m=10k:N$, and $q=1.8$. Fig. \ref{fig:5} shows that for all the tested cases, the $q$-ratio based bounds are smaller than those based on the RIC. For some certain $m$ and $k$, the $q$-ratio CMSV based bounds apply even when the RIC based bound do not apply (i.e., $\delta_{2k}(A)\geq \sqrt{2}-1$). When $m$ approaches $N$, it can be observed that the $q$-ratio based bounds are slightly larger than $2$  while the RIC based bounds approach $4k^{1/1.8-1/2}\geq 4$ as $\delta_{2k}(A)\rightarrow 0$. 
	
	\section{Conclusion}
	In this paper, we proposed a new measure of the measurement matrix's incoherence, the $q$-ratio CMSV which was defined based on the $q$-ratio sparsity measure. We established the bounds for both  $\ell_q$ norm and $\ell_1$ norm of the reconstruction errors of the Basis Pursuit, the Dantzig selector and the Lasso estimator using the $q$-ratio CMSV. For the subgaussian random matrices, we showed that the $q$-ratio CMSV is bounded away from zero as long as the number of measurements is relatively large. A CCP algorithm was developed to verify the sufficient conditions guaranteeing the unique noiseless recovery and an interior point problem was used to compute the $q$-ratio CMSV. Numerical experiments were presented to illustrate our theoretical results and assess all the algorithms. Some further generalizations including the block sparsity recovery, low-rank matrix recovery or more general high dimensional M-estimation are left for future work.

	
	%

	\appendix[Proofs.]
	
	\noindent
	\emph{Proof of Proposition 1.} If there exists $z\in\mathrm{ker} A\setminus \{0\}$ and $|S|\leq k$ such that $\lVert z_S\rVert_1\geq \lVert z_{S^c}\rVert_1$, then for any $1<q\leq \infty$, we have  
	\begin{align*}
	\lVert z\rVert_1=\lVert z_S\rVert_1+\lVert z_{S^c}\rVert_1\leq 2\lVert z_S\rVert_1  
	&\leq 2k^{1-1/q}\lVert z_S\rVert_q \\
	&\leq 2k^{1-1/q}\lVert z\rVert_q,
	\end{align*}
	which implies that $k\geq 2^{\frac{q}{1-q}} s_q(x)$ for any $1<q\leq\infty$.
	
	As a consequence of contraposition, when there exists some $1<q\leq \infty$ such that $k<\min\limits_{z\in\mathrm{ker} A\setminus\{0\}}\,\,2^{\frac{q}{1-q}}s_q(z)$, $\lVert z_S\rVert_1<\lVert z_{S^c}\rVert_1$ holds that for all $z\in\mathrm{ker} A\setminus \{0\}$ and $|S|\leq k$. Thus the null space property of order $k$ is fulfilled and the unique solution to problem (\ref{nobp}) is exactly the true $k$-sparse signal $x$.\\
	
	\noindent
	\emph{Proof of Proposition 2.} We firstly prove the left hand side of (\ref{rhoineq}). For any $z\in\mathbb{R}^N\setminus\{0\}$ and $1<q_2\leq q_1\leq \infty$, when $s_{q_1}(z)\leq s$, we have $\left(\frac{\lVert z\rVert_1}{\lVert z\rVert_{q_1}}\right)^{\frac{q_1}{q_1-1}}\leq s\Rightarrow \lVert z\rVert_1\leq s^{\frac{q_1-1}{q_1}}\lVert z\rVert_{q_1}\leq s^{\frac{q_1-1}{q_1}}\lVert z\rVert_{q_2}$ as $\lVert z\rVert_{q_2}\geq \lVert z\rVert_{q_1}$. Then, we have $$
	\frac{\lVert z\rVert_1}{\lVert z\rVert_{q_2}}\leq s^{\frac{q_1-1}{q_1}}\Rightarrow s_{q_2}(z)=\left(\frac{\lVert z\rVert_1}{\lVert z\rVert_{q_2}}\right)^{\frac{q_2}{q_2-1}}\leq s^{\frac{q_2(q_1-1)}{q_1(q_2-1)}},
	$$
	which implies that $$
	\{z: s_{q_1}(z)\leq s\}\subseteq \{z: s_{q_2}(z)\leq s^{\frac{q_2(q_1-1)}{q_1(q_2-1)}}\}.
	$$
	Therefore, we have \begin{align*}
	\rho_{q_1,s}(A)&=\min\limits_{z\neq 0,s_{q_1}(z)\leq s}\frac{\lVert Az\rVert_2}{\lVert z\rVert_{q_1}}\geq \min\limits_{z\neq 0, s_{q_2}(z)\leq s^{\frac{q_2(q_1-1)}{q_1(q_2-1)}}}\frac{\lVert Az\rVert_2}{\lVert z\rVert_{q_1}} \\
	&= \min\limits_{z\neq 0, s_{q_2}(z)\leq s^{\frac{q_2(q_1-1)}{q_1(q_2-1)}}} \frac{\lVert Az\rVert_2}{\lVert z\rVert_{q_2}}\cdot\frac{\lVert z\rVert_{q_2}}{\lVert z\rVert_{q_1}} \\
	&\geq  \min\limits_{z\neq 0, s_{q_2}(z)\leq s^{\frac{q_2(q_1-1)}{q_1(q_2-1)}}} \frac{\lVert Az\rVert_2}{\lVert z\rVert_{q_2}}=\rho_{q_2,s^{\frac{q_2(q_1-1)}{q_1(q_2-1)}}}(A).
	\end{align*}
	
	Next we verify the right hand side of (\ref{rhoineq}). For any $z\in\mathbb{R}^N\setminus\{0\}$, by using the non-increasing property of the $q$-ratio sparsity, we have $\frac{\lVert z\rVert_{1}}{\lVert z\rVert_{\infty}}=s_\infty(z)\leq s_{q_2}(z)\leq s^{\frac{q_2(q_1-1)}{q_1(q_2-1)}}$ since $q_2\leq \infty$. Then as $1<q_2\leq q_1\leq \infty$, thus it holds that $\frac{\lVert z\rVert_{q_2}}{\lVert z\rVert_{q_1}}\leq \frac{\lVert z\rVert_{1}}{\lVert z\rVert_{\infty}}\leq s^{\frac{q_2(q_1-1)}{q_1(q_2-1)}}\Rightarrow \lVert z\rVert_{q_2}< s^{\frac{q_2(q_1-1)}{q_1(q_2-1)}}\lVert z\rVert_{q_1}$. In addition, the non-increasing property of $q$-ratio sparsity $s_{q_1}(z)\leq s_{q_2}(z)$ implies that 
	$$
	\{z: s_{q_2}(z)\leq s^{\frac{q_2(q_1-1)}{q_1(q_2-1)}}\}\subseteq \{z: s_{q_1}(z)\leq s^{\frac{q_2(q_1-1)}{q_1(q_2-1)}} \}.
	$$
	Therefore, we have \begin{align*}
	\rho_{q_2,s^{\frac{q_2(q_1-1)}{q_1(q_2-1)}}}(A)&=\min\limits_{z\neq 0,s_{q_2}(z)\leq s^{\frac{q_2(q_1-1)}{q_1(q_2-1)}}}\frac{\lVert Az\rVert_2}{\lVert z\rVert_{q_2}} \\
	&\geq \min\limits_{z\neq 0, s_{q_1}(z)\leq  s^{\frac{q_2(q_1-1)}{q_1(q_2-1)}}} \frac{\lVert Az\rVert_2}{\lVert z\rVert_{q_2}} \\
	&= \min\limits_{z\neq 0, s_{q_1}(z)\leq  s^{\frac{q_2(q_1-1)}{q_1(q_2-1)}}} \frac{\lVert Az\rVert_2}{\lVert z\rVert_{q_1}}\cdot\frac{\lVert z\rVert_{q_1}}{\lVert z\rVert_{q_2}}\\
	&\geq s^{-\frac{q_2(q_1-1)}{q_1(q_2-1)}} \min\limits_{z\neq 0,s_{q_1}(z)\leq s^{\frac{q_2(q_1-1)}{q_1(q_2-1)}}} \frac{\lVert Az\rVert_2}{\lVert z\rVert_{q_1}} \\
	&=s^{-\frac{q_2(q_1-1)}{q_1(q_2-1)}} \rho_{q_1, s^{\frac{q_2(q_1-1)}{q_1(q_2-1)}}}(A).
	\end{align*}
	The proof is completed. \\

	\noindent 
	\emph{Proof of Theorem 1.} The proof procedure follows from the similar arguments in \cite{tn1,tn3}, which is simpler than those employed for obtaining the RIC based bounds. The derivation has two key steps: \\
	
	\emph{Step 1}: For all algorithms, show that the residual $h=\hat{x}-x$ is $q$-ratio sparse. As $x$ is $k$-sparse, we assume that $\mathrm{supp}(x)=S$ and $|S|\leq k$. \\
	
	First, for BP and DS, since $\lVert \hat{x}\rVert_1=\lVert x+h\rVert_1$ is the minimum among all $z$ satisfying the constraints of BP and DS (including the true signal $x$), we have \begin{align*}
	\lVert x\rVert_1&\geq \lVert \hat{x}\rVert_1=\lVert x+h\rVert_1=\lVert x_S+h_S\rVert_1+\lVert x_{S^c}+h_{S^c}\rVert_1 \\
	&\geq \lVert x_S\rVert_1-\lVert h_S\rVert_1+\lVert h_{S^c}\rVert_1 \\
	&=\lVert x\rVert_1-\lVert h_S\rVert_1+\lVert h_{S^c}\rVert_1.
	\end{align*}
	Therefore, we obtain that $\lVert h_{S^c}\rVert_1\leq \lVert h_{S}\rVert_1$, which leads to  \begin{align*}
	\lVert h\rVert_1&=\lVert h_S\rVert_1+\lVert h_{S^c}\rVert_1 \\
	&\leq 2\lVert h_S\rVert_1\leq 2k^{1-1/q}\lVert h_S\rVert_{q}\leq 2k^{1-1/q}\lVert h\rVert_{q},
	\end{align*}
	for any $1<p\leq \infty$. Thus, $s_{q}(h)=\left(\frac{\lVert h\rVert_1}{\lVert h\rVert_q}\right)^{\frac{q}{q-1}}\leq 2^{\frac{q}{q-1}} k$. 
	
	Next, as for Lasso, since the noise $w$ satisfies $\lVert A^T w\rVert_\infty\leq \kappa \lambda_N \sigma$ for some small $\kappa>0$ and $\hat{x}$ is a solution of Lasso, we have $$
	\frac{1}{2}\lVert A\hat{x}-y\rVert_2^2+\lambda_N\sigma\lVert \hat{x}\rVert_1\leq \frac{1}{2}\lVert Ax-y\rVert_2^2+\lambda_N\sigma\lVert x\rVert_1.
	$$
	Consequently, substituting $y=Ax+w$ yields \begin{align*}
	\lambda_N\sigma\lVert\hat{x}\rVert_1&\leq \frac{1}{2}\lVert w\rVert_2^2-\frac{1}{2}\lVert A(\hat{x}-x)-w\rVert_2^2+\lambda_N\sigma\lVert x\rVert_1\\
	&\leq \frac{1}{2}\lVert w\rVert_2^2-\frac{1}{2}\lVert A(\hat{x}-x)\rVert_2^2+\langle A(\hat{x}-x),w\rangle \\
	&\quad -\frac{1}{2}\lVert w\rVert_2^2+\lambda_N\sigma\lVert x\rVert_1\\
	&\leq \langle A(\hat{x}-x),w\rangle+\lambda_N\sigma\lVert x\rVert_1 \\
	&=\langle \hat{x}-x, A^Tw\rangle+\lambda_N\sigma\lVert x\rVert_1 \\
	&\leq \lVert \hat{x}-x\rVert_1\lVert A^T w\rVert_{\infty}+\lambda_N\sigma\lVert x\rVert_1 \\
	&\leq \kappa \lambda_N\sigma\lVert h\rVert_1+\lambda_N\sigma\lVert x\rVert_1,
	\end{align*}
	which leads to \begin{align}
	\lVert \hat{x}\rVert_1\leq \kappa\lVert h\rVert_1+\lVert x\rVert_1. \label{lasso}
	\end{align}
	Therefore, it holds that \begin{align*}
	\lVert x\rVert_1&\geq \lVert \hat{x}\rVert_1-\kappa \lVert h\rVert_1\\
	&=\lVert x+h_{S^c}+h_S\rVert_1-\kappa\lVert h_{S^c}+h_S\rVert_1 \\
	&\geq \lVert x+h_{S^c}\rVert_1-\lVert h_S\rVert_1-\kappa(\lVert h_{S^c}\rVert_1+\lVert h_{S}\rVert_1)\\
	&=\lVert x\rVert_1+(1-\kappa)\lVert h_{S^c}\rVert_1-(1+\kappa)\lVert h_S\rVert_1.
	\end{align*}
	As a consequence, we obtain $$
	\lVert h_{S^c}\rVert_1\leq \frac{1+\kappa}{1-\kappa}\lVert h_S\rVert_1,
	$$
	which implies that \begin{align*}
	\lVert h\rVert_1=\lVert h_{S^c}\rVert_1+\lVert h_S\rVert_1&\leq \frac{2}{1-\kappa}\lVert h_S\rVert_1\\
	&\leq \frac{2}{1-\kappa}k^{1-1/q}\lVert h_S\rVert_q   \\
	&\leq \frac{2}{1-\kappa}k^{1-1/q}\lVert h\rVert_q.
	\end{align*}
	Thus, it yields $$
	s_q(h)=\left(\frac{\lVert h\rVert_1}{\lVert h\rVert_q}\right)^{\frac{q}{q-1}}\leq \left(\frac{2}{1-\kappa}\right)^{\frac{q}{q-1}}k.
	$$\\
	
	\emph{Step 2.} Obtain an upper bound on $\lVert Ah\rVert_2$ and then get the $\ell_q$ norm and $\ell_1$ norm bounds on the error vector $h$ via the definition of $q$-ratio CMSV. \\
	
	(a) For the BP, this is trivial since both $x$ and $\hat{x}$ satisfy the constraint $\lVert y-Az\rVert_2\leq \varepsilon$, the triangle inequality implies \begin{align}
	\lVert Ah\rVert_2=\lVert A(\hat{x}-x)\rVert_2&\leq \lVert A\hat{x}-y\rVert_2+\lVert y-Ax\rVert_2 \nonumber \\
	&\leq 2\varepsilon. \label{ahbp}
	\end{align}
	Then it follows from the definition of $q$-ratio CMSV and $s_q(h)\leq 2^{\frac{q}{q-1}}k$ that $$
	\rho_{q,2^{\frac{q}{q-1}}k}(A)\lVert h\rVert_q\leq \lVert Ah\rVert_2\leq 2\varepsilon\Rightarrow \lVert h\rVert_q\leq \frac{2\varepsilon}{\rho_{q,2^{\frac{q}{q-1}}k}(A)}.
	$$
	Meanwhile, $\lVert h\rVert_1\leq 2k^{1-1/q}\lVert h\rVert_{q}\Rightarrow \lVert h\rVert_1\leq \frac{4k^{1-1/q}\varepsilon}{\rho_{q,2^{\frac{q}{q-1}}k}(A)}$. \\
	
	(b) Now for the DS, since $\lVert A^Tw\rVert_{\infty}\leq \lambda_N\sigma$, \begin{align*}
	\lVert A^T Ah\rVert_{\infty}\leq \lVert A^T(y-A\hat{x})\rVert_{\infty}+\lVert A^T(y-Ax)\rVert_{\infty} \leq 2\lambda_N\sigma.
	\end{align*}
	Therefore, we have \begin{align}
	\lVert Ah\rVert_2^2=h^TA^TAh&=\sum\limits_{i=1}^N h_i(A^TAh)_i\leq \sum\limits_{i=1}^N|h_i||(A^TAh)_i| \nonumber \\
	&\leq \lVert A^TAh\rVert_{\infty}\lVert h\rVert_1\leq 2\lambda_N\sigma\lVert h\rVert_1. \label{ahds}
	\end{align}
	Thus, with $s_q(h)\leq 2^{\frac{q}{q-1}}k$, \begin{align*}
	&\rho_{q,2^{\frac{q}{q-1}}k}^2(A)\lVert h\rVert_{q}^2\leq \lVert Ah\rVert_2^2\leq 2\lambda_N\sigma\lVert h\rVert_1\leq 4\lambda_N\sigma k^{1-1/q}\lVert h\rVert_q \\
	&\Rightarrow \lVert h\rVert_q\leq \frac{4k^{1-1/q}}{\rho_{q,2^{\frac{q}{q-1}}k}^2(A)}\lambda_N\sigma.
	\end{align*}
	Hence, $\lVert h\rVert_1\leq 2k^{1-1/q}\lVert h\rVert_{q}\leq \frac{8k^{2-2/q}}{\rho_{q,2^{\frac{q}{q-1}}k}^2(A)}\lambda_N\sigma$. \\
	
	(c) Finally, we establish an upper bound on $\lVert Ah\rVert_2^2$ for the Lasso with $\lVert A^Tw\rVert_\infty\leq \kappa \lambda_N\sigma$. \begin{align*}
	\lVert A^TAh\rVert_{\infty}&\leq \lVert A^T(y-Ax)\rVert_{\infty}+\lVert A^T(y-A\hat{x})\rVert_{\infty} \\
	&\leq \lVert A^Tw\rVert_\infty +\lVert A^T(y-A\hat{x})\rVert_\infty \\
	&\leq \kappa\lambda_N\sigma+\lVert A^T(y-A\hat{x})\rVert_{\infty}.
	\end{align*}
	Moreover, since $\hat{x}$ is the solution of Lasso, the optimality condition yields that $$
	A^T(y-A\hat{x})\in\lambda_N\sigma\partial \lVert \hat{x}\rVert_1,
	$$
	where $\partial \lVert \hat{x}\rVert_1=[-1,1]^N$ is the subgradient of $\lVert \cdot\rVert_1$ evaluated at $\hat{x}$. Thus, we have $\lVert A^T(y-A\hat{x})\rVert_\infty\leq \lambda_N\sigma$, which leads to $$
	\lVert A^TAh\rVert_\infty\leq (\kappa+1)\lambda_N\sigma.
	$$
	Following the same argument as before, we get \begin{align}
	\lVert Ah\rVert_2^2\leq (\kappa+1)\lambda_N\sigma\lVert h\rVert_1. \label{ahlasso}
	\end{align}
	As a consequence, with $s_q(h)\leq \left(\frac{2}{1-\kappa}\right)^{\frac{q}{q-1}}k$, \begin{align}
	\rho_{q,(\frac{2}{1-\kappa})^{\frac{q}{q-1}}k}^2(A)\lVert h\rVert_q^2&\leq \lVert Ah\rVert_2^2\leq (\kappa+1)\lambda_N\sigma\lVert h\rVert_1 \nonumber \\
	&\leq \lambda_N\sigma\frac{2(\kappa+1)}{1-\kappa}k^{1-1/q}\lVert h\rVert_q,
	\end{align}
	which implies that $$
	\lVert h\rVert_q\leq \frac{k^{1-1/q}}{\rho_{q,\left(\frac{2}{1-\kappa}\right)^{\frac{q}{q-1}}k}^2(A)}\cdot \frac{2(\kappa+1)}{1-\kappa}\lambda_N\sigma.
	$$
	Therefore, (\ref{sparselassol1}) holds since $\lVert h\rVert_1\leq \frac{2}{1-\kappa}k^{1-1/q}\lVert h\rVert_q$. \\
	
	\noindent\\
	\emph{Proof of Theorem 2.}  Assume that $S$ is the index set that contains the largest $k$ absolute entries of $x$ so that $\sigma_k(x)_1=\lVert x_{S^c}\rVert_1$ and let $h=\hat{x}-x$. The derivations also have two steps: \\
	
	\emph{Step 1}: For all algorithms, bound $\lVert h\rVert_1$ with $\lVert h\rVert_q$ and $\lVert x_{S^c}\rVert_1$. \\
	
	First for BP and DS, since $\lVert \hat{x}\rVert_1=\lVert x+h\rVert_1$ is the minimum among all $z$ satisfying the constrains of BP and DS, we have \begin{align*}
	\lVert x_S\rVert_1+\lVert x_{S^c}\rVert_1&=\lVert x\rVert_1\geq \lVert \hat{x}\rVert_1=\lVert x+h\rVert_1 \\
	&=\lVert x_S+h_S\rVert_1+\lVert x_{S^c}+h_{S^c}\rVert_1\\
	&\geq \lVert x_S\rVert_1-\lVert h_S\rVert_1-\lVert x_{S^c}\rVert_1+\lVert h_{S^c}\rVert_1, \label{error}
	\end{align*}
	which implies that \begin{align}
	\lVert h_{S^c}\rVert_1\leq \lVert h_S\rVert_1+2\lVert x_{S^c}\rVert_1.
	\end{align}
	As a consequence, \begin{align}
	\lVert h\rVert_1=\lVert h_S\rVert_1+\lVert h_{S^c}\rVert_1&\leq 2\lVert h_S\rVert_1+2\lVert x_{S^c}\rVert_1  \nonumber \\
	&\leq 2k^{1-1/q}\lVert h_S\rVert_q+2\lVert x_{S^c}\rVert_1  \nonumber \\
	&\leq 2k^{1-1/q}\lVert h\rVert_q+2\lVert x_{S^c}\rVert_1. \label{errorbp} 
	\end{align}
	
	Next, regarding to Lasso, adopting (\ref{lasso}), we obtain that \begin{align*}
	&\lVert x_S\rVert_1+\lVert x_{S^c}\rVert_1=\lVert x\rVert_1\geq \lVert \hat{x}\rVert_1-\kappa\lVert h\rVert_1 \\
	&\quad\geq \lVert x_S+x_{S^c}+h_S+h_{S^c}\rVert_1-\kappa\lVert h_S+h_{S^c}\rVert_1 \\
	&\quad\geq \lVert x_S+h_{S^c}\rVert_1-\lVert x_{S^c}\rVert_1-\lVert h_S\rVert_1-\kappa\lVert h_S\rVert_1-\kappa\lVert h_{S^c}\rVert_1 \\
	&\quad=\lVert x_S\rVert_1+(1-\kappa)\lVert h_{S^c}\rVert_1-\lVert x_{S^c}\rVert_1-(1+\kappa)\lVert h_S\rVert_1,
	\end{align*}
	which implies that \begin{align}
	\lVert h_{S^c}\rVert_1\leq \frac{1+\kappa}{1-\kappa}\lVert h_S\rVert_1+\frac{2}{1-\kappa}\lVert x_{S^c}\rVert_1.
	\end{align}
	Therefore, we have \begin{align}
	\lVert h\rVert_1&\leq \lVert h_S\rVert_1+\lVert h_{S^c}\rVert_1 \nonumber \\
	&\leq \frac{2}{1-\kappa}\lVert h_S\rVert_1+\frac{2}{1-\kappa}\lVert x_{S^c}\rVert_1 \nonumber \\
	&\leq  \frac{2}{1-\kappa}k^{1-1/q}\lVert h\rVert_q+\frac{2}{1-\kappa}\lVert x_{S^c}\rVert_1. \label{errorlasso}
	\end{align}\\
	
	\emph{Step 2}: Verify that the $q$-ratio sparsity levels of $h$ have lower bounds if $\lVert h\rVert_q$ is larger than the bound component caused by the error.\\
	
	(a) Specifically, for the BP, we assume that $h\neq 0$ and $\lVert h\rVert_q>\frac{2\varepsilon}{\rho_{q,4^{\frac{q}{q-1}}k}(A)}$, otherwise (\ref{robust1}) holds trivially. Since $\lVert Ah\rVert_2\leq 2\varepsilon$, see (\ref{ahbp}), so we have $\lVert h\rVert_q>\frac{\lVert Ah\rVert_2}{\rho_{q,4^{\frac{q}{q-1}}k}(A)}$. Then it holds that \begin{align}
	\frac{\lVert Ah\rVert_2}{\lVert h\rVert_q}<{\rho_{q,4^{\frac{q}{q-1}}k}(A)}=\min\limits_{z\neq 0, s_q(z)\leq 4^{\frac{q}{q-1}}k}\frac{\lVert Az\rVert_2}{\lVert z\rVert_q} \nonumber \\
	\Rightarrow s_q(h)>4^{\frac{q}{q-1}}k\Rightarrow \lVert h\rVert_1>4k^{1-1/q}\lVert h\rVert_q.
	\end{align}
	Combining (\ref{errorbp}), we have $\lVert h\rVert_q<k^{1/q-1}\lVert x_{S^c}\rVert_1=k^{1/q-1}\sigma_{k}(x)_1$, which completes the proof of (\ref{robust1}). The error $\ell_1$ norm bound (\ref{robust1l1}) follows immediately from (\ref{robust1}) and (\ref{errorbp}).\\
	
	(b) As for the DS, we assume $h\neq 0$ and $\lVert h\rVert_q>\frac{8k^{1-1/q}}{\rho_{q,4^{\frac{q}{q-1}}k}^2(A)}\lambda_N\sigma$, otherwise (\ref{robust2}) holds trivially. As $\lVert Ah\rVert_2^2\leq 2\lambda_N\sigma\lVert h\rVert_1$, see (\ref{ahds}), so we have $\lVert h\rVert_q>\frac{4k^{1-1/q}}{\rho_{q,4^{\frac{q}{q-1}}k}^2(A)}\cdot \frac{\lVert Ah\rVert_2^2}{\lVert h\rVert_1}$. Then it implies that \begin{align}
	&\rho_{q,4^{\frac{q}{q-1}}k}^2(A)=\min\limits_{z\neq 0, s_q(z)\leq 4^{\frac{q}{q-1}}k}\frac{\lVert Az\rVert_2^2}{\lVert z\rVert_q^2} \nonumber \\
	&\qquad\qquad>\frac{\lVert Ah\rVert_2^2}{\lVert h\rVert_q^2}\left(\frac{4^{\frac{q}{q-1}}k}{s_q(h)}\right)^{1-1/q} \nonumber \\
	&\Rightarrow s_q(h)>4^{\frac{q}{q-1}}k\Rightarrow \lVert h\rVert_1>4k^{1-1/q}\lVert h\rVert_q.
	\end{align}
	Combining (\ref{errorbp}), we have $\lVert h\rVert_q<k^{1/q-1}\lVert x_{S^c}\rVert_1=k^{1/q-1}\sigma_{k}(x)_1$, which completes the proof of (\ref{robust2}). Then (\ref{robust2l1}) holds as a result of (\ref{robust2}) and (\ref{errorbp}).\\ 
	
	(c) Finally, regarding to the Lasso, we assume that $h\neq 0$ and $\lVert h\rVert_q>\frac{1+\kappa}{1-\kappa}\cdot\frac{4k^{1-1/q}}{\rho_{q,(\frac{4}{1-\kappa})^{\frac{q}{q-1}}k}^2(A)}\lambda_N\sigma$, otherwise (\ref{robust3}) holds trivially. Since in this case $\lVert Ah\rVert_2^2\leq (1+\kappa)\lambda_N\sigma\lVert h\rVert_1$, see (\ref{ahlasso}), so we have $\lVert h\rVert_q>\frac{4k^{1-1/q}}{(1-\kappa)\rho_{q,(\frac{4}{1-\kappa})^{\frac{q}{q-1}}k}^2(A)}\cdot\frac{\lVert Ah\rVert_2^2}{\lVert h\rVert_1}$. Then it leads to \begin{align}
	\rho_{q,(\frac{4}{1-\kappa})^{\frac{q}{q-1}}k}^2(A)&=\min\limits_{z\neq 0, s_q(z)\leq (\frac{4}{1-\kappa})^{\frac{q}{q-1}}k}\frac{\lVert Az\rVert_2^2}{\lVert z\rVert_q^2} \nonumber \\
	&>\frac{\lVert Ah\rVert_2^2}{\lVert h\rVert_q^2}\left(\frac{(\frac{4}{1-\kappa})^{\frac{q}{q-1}}k}{s_q(h)}\right)^{1-\frac{1}{q}} \nonumber \\
	&\Rightarrow s_q(h)>(\frac{4}{1-\kappa})^{\frac{q}{q-1}}k \nonumber \\
	&\Rightarrow \lVert h\rVert_1>\frac{4}{1-\kappa}k^{1-1/q}\lVert h\rVert_q.
	\end{align}
	Combining (\ref{errorlasso}), we have $\lVert h\rVert_q<k^{1/q-1}\lVert x_{S^c}\rVert_1=k^{1/q-1}\sigma_{k}(x)_1$, which completes the proof of (\ref{robust3}). Consequently, (\ref{robust3l1}) is obtained by (\ref{robust3}) and (\ref{errorlasso}).\\
	
	\section*{Acknowledgment}
	
	This work is supported by the Swedish Research Council grant (Reg.No. 340-2013-5342).

	\ifCLASSOPTIONcaptionsoff
	\newpage
	\fi

	
	
	%
	\bibliographystyle{IEEEtran}
\end{document}